\documentclass[sigconf,10pt]{acmart}

\AtBeginDocument{%
  \providecommand\BibTeX{{%
    \normalfont B\kern-0.5em{\scshape i\kern-0.25em b}\kern-0.8em\TeX}}}

\setcopyright{none}
\renewcommand\footnotetextcopyrightpermission[1]{} 
\usepackage[english]{babel}
\usepackage{graphicx}
\usepackage{graphics}
\usepackage[utf8]{inputenc}
\usepackage{amsmath}
\usepackage{tabularx}
\usepackage{array, multirow} 
\usepackage{subcaption}
\usepackage{hyperref} 
\usepackage{xurl}    
\usepackage{listings}
\usepackage[normalem]{ulem}
\usepackage{enumitem}
\usepackage{tikz}

\usepackage{adjustbox}

\usetikzlibrary{positioning,fit,arrows.meta,calc,backgrounds}

\newboolean{showcomments}
\setboolean{showcomments}{false}

\newcommand{\sys}{{\em BQT+ }}

\usepackage{pifont}

\begin{document}

\title{Robust and Extensible Measurement of Broadband Plans with \sys}

\author{
    Laasya Koduru\textsuperscript{*}, 
    Sylee (Roman) Beltiukov\textsuperscript{*}, 
    Alexander Nguyen\textsuperscript{*}, 
    Eugene Vuong\textsuperscript{\dag}, 
    Jaber Daneshamooz\textsuperscript{*},
    Tejas Narechania\textsuperscript{\ddag}, 
    Elizabeth Belding\textsuperscript{*},
    Arpit Gupta\textsuperscript{*}
}

\affiliation{
  \institution{
    \quad \textsuperscript{*} University of California Santa Barbara 
    \quad \textsuperscript{\dag} California State University East Bay \\
    \quad \textsuperscript{\ddag} University of California Berkeley
    \country{} 
}
}

\renewcommand{\shortauthors}{Laasya Koduru, Sylee Beltiukov, Alexander Nguyen, Eugene Vuong, Jaber Daneshamooz, Tejas Narechania, Elizabeth Belding, Arpit Gupta}
\newcommand{\smartparagraph}[1]{\noindent{\bf #1}\ }

\begin{sloppypar}
\begin{abstract}

Independent, street address-level broadband data is essential for evaluating  Internet infrastructure investments, such as the \$42B Broadband Equity, Access, and Deployment (BEAD) program. Evaluating these investments requires longitudinal visibility into broadband availability, quality, and affordability, including data on pre-disbursement baselines and changes in providers' advertised plans. While such data can be obtained through Internet Service Provider (ISP) web interfaces, these workloads impose three fundamental system requirements: robustness to frequent interface evolution, extensibility across hundreds of providers, and low technical overhead for non-expert users. Existing systems fail to meet these three essential requirements. %
We present {\em BQT+}, a broadband plan measurement framework that replaces monolithic workflows with declarative state/action specifications. \sys models querying intent as an interaction state space, formalized as an abstract nondeterministic finite automaton (NFA), and selects execution paths at runtime to accommodate alternative interaction flows and localized interface changes. We show that \sys sustains longitudinal monitoring of 64 ISPs, supporting querying for over 100 ISPs. We apply it to two policy studies: constructing a BEAD pre-disbursement baseline and benchmarking broadband affordability across over 124{,}000 addresses in four states.

\end{abstract}

\maketitle
\thispagestyle{plain}

\section{Introduction}
\label{sec:intro}

Effective broadband policymaking requires accurate, fine-grained data that characterizes the current and historical state of broadband availability, quality, and affordability. This need arises across multiple, concurrent policy workloads, particularly as the United States (U.S.) disburses over \$42B through the Broadband Equity, Access, and Deployment (BEAD) program. Policymakers require this data not just as a snapshot, but longitudinally, to evaluate the effectiveness of past investments and guide current initiatives. 
Unfortunately, existing regulatory broadband datasets are not designed to satisfy these requirements. Historical mechanisms~\cite{fcc-form477} and recent efforts~\cite{fcc-bdc,fcc-nbm} rely on provider self-reporting and have been shown to overstate coverage~\cite{form477-inaccuracy, nbm-overstating-coverage}.

\begin{figure*}[t]
    \centering
    \includegraphics[width=.9\linewidth]{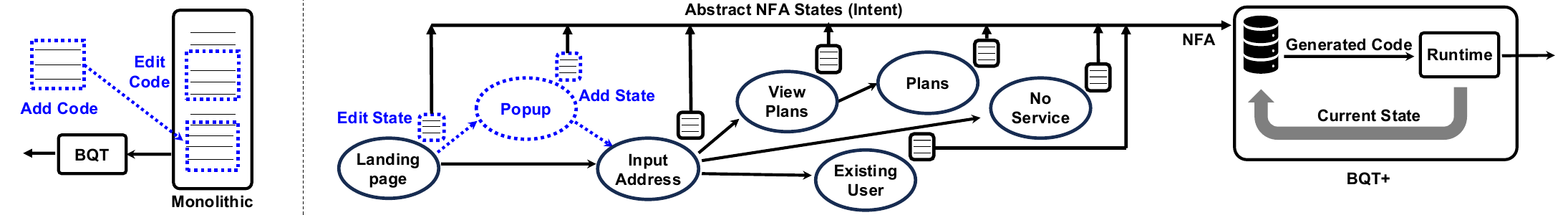}
    \caption{BQT vs. \sys. BQT encodes each ISP BAT as a single deterministic workflow in monolithic code; \sys encodes intent as an interaction state space (abstract NFA) traversed at runtime. Dotted/blue markings show where interface evolution forces changes: BQT edits/adds workflow code, while \sys adds/updates individual state specifications, and traverses from the currently observed state.}
    \label{fig:BQT-vs-BQT+-system}
\end{figure*}

Internet Service Providers (ISPs), however, necessarily expose address-specific service availability, speed tiers, and prices through consumer-facing service qualification websites, which prospective customers use to determine what service can actually be purchased at a given location. These interfaces---commonly referred to as Broadband Availability Tools (BATs)~\cite{bat}---therefore provide a direct, consumer-visible view of broadband plans.
Independently querying these interfaces enables verifiable, policy-grade, address-level broadband measurement data on availability, advertised quality, and pricing.

\smartparagraph{Implications for systems design.}
Requiring independent querying of ISP interfaces at policy scale fundamentally changes the nature of the measurement problem. Unlike targeted or one-off studies, policy workloads such as BEAD evaluation and affordability monitoring demand repeated, longitudinal measurement across hundreds of national, regional, and local ISPs, whose BATs evolve frequently and independently. 

Supporting such an operation imposes three system requirements. First, \emph{extensibility}: the ability to incorporate and operate across a large and heterogeneous set of ISPs without incurring prohibitive per-provider development cost. Second, \emph{robustness}: the ability to sustain long-running operation despite frequent and unsynchronized changes to ISP interfaces~\cite{ui-drift-web}. Third, \emph{low technical overhead}: the ability to specify, operate, and maintain querying specifications without extensive low-level engineering effort. This requirement is essential because broadband policy analysis is conducted not only by systems researchers, but also by policymakers, regulatory agencies, legal scholars, social scientists, nonprofit organizations, and legislative staff, who rely on broadband data but typically lack deep computing expertise. For policy-grade measurement to be sustainable in practice, querying systems must therefore minimize technical debt and reduce dependence on specialized engineering interventions.

\smartparagraph{Limitations of existing querying systems.}
Prior work enabled independent broadband plan measurement through the Broadband-plan Querying Tool (BQT)~\cite{bqt}, which emulates consumer interactions with ISP websites to extract address-level availability, advertised speed tiers, and pricing, and has supported influential policy analyses~\cite{decoding-the-divide, caf-study}. As shown in Figure~\ref{fig:BQT-vs-BQT+-system}, BQT encodes each ISP interaction as a deterministic, end-to-end workflow in monolithic low-level execution code, prescribing a complete sequence of actions from address input to plan discovery.

The limitation of BQT is rooted in the abstraction: workflow-centric designs conflate interaction structure with execution order in a single artifact, implicitly assuming one stable execution trace. In practice, ISP BATs evolve and often admit multiple valid paths (e.g., inserted intermediate steps such as popups, reordered interactions, or minor changes in existing workflow), so changes that do not affect measurement intent still require direct, dispersed edits (marked blue in Figure~\ref{fig:BQT-vs-BQT+-system}) across imperative control flow that are difficult to localize or reuse across interface revisions. This design also does not naturally represent partial workflows or alternate control flow without invasive code changes. At the policy scale, these properties accumulate technical debt, yield engineering effort that scales with the number of ISPs, complicate sustained longitudinal operation, and limit practical use to stakeholders without substantial technical expertise. These limitations motivate an abstraction that represents interaction structure independently of execution behavior and treats alternative paths as first-class elements.

\smartparagraph{Proposed solution.}
We address this abstraction gap by introducing \textbf{\emph{BQT+}}, which models ISP consumer-facing interfaces as interaction state spaces. This abstraction induces two complementary forms of disaggregation. Figure~\ref{fig:BQT-vs-BQT+-system} illustrates this design. First, \sys decomposes monolithic workflows into \emph{state specifications}: querying intent describes observable interaction states and terminal outcomes that define successful plan discovery, without prescribing a complete navigation sequence. This formulation simplifies intent specification and promotes reuse across interface variants. In Figure~\ref{fig:BQT-vs-BQT+-system}, blue lines represent updates to existing states, while dotted lines represent added states, enabling reuse under ISP interface updates. Second, \sys separates execution from intent by treating execution as a runtime traversal policy over the interaction state space. By expressing intent as declarative state/action specifications rather than low-level imperative workflow code, \sys localizes maintenance under interface churn and reduces technical debt, lowering the barrier to authoring and upkeep for non-expert stakeholders.

We formalize this abstraction using a \textit{\textbf{nondeterministic finite automaton (NFA)}}~\cite{finite-automata}, where states are detector-defined interface conditions over observable cues and transitions map state/action pairs to a set of possible successor states. The NFA representation supports partial workflows, enables dynamic branching, and resolves execution decisions at runtime. As interfaces evolve, execution adapts by selecting alternate traversal paths over an incrementally updated interaction space, allowing \sys to support long-running operations across heterogeneous ISPs while making robustness and extensibility intrinsic properties of the abstraction.

\smartparagraph{Contributions.}
This paper makes three key contributions:

\begin{itemize}[leftmargin=*,nosep]
    \item {\bf An interaction-state abstraction.} We introduce an abstraction that represents querying intent independently of execution, modeling interfaces as an interaction state space formalized as an abstract NFA over detector-defined states and admissible actions.

    \item {\bf \sys: a broadband-plan querying system.}
    We design and implement \emph{BQT+}, which replaces monolithic imperative workflows with declarative state specifications and executes queries via runtime traversal of the interaction state space. This design improves \emph{extensibility} through decomposed, declarative specifications and reuse across providers; improves \emph{robustness} through localized updates under interface churn; and reduces \emph{technical debt} by lowering the barriers to authoring and upkeep for non-expert users.

    \item {\bf Policy applications.} We apply \sys to two policy workloads: constructing a BEAD pre-disbursement baseline in four states and benchmarking broadband affordability at 124{,}000 addresses, including evidence used in a statewide affordability study presented to legislators in Virginia.
\end{itemize}

\section{Background and Motivation}
\label{sec:background}

\subsection{Data Requirements}
\label{ssec:policy-workloads}

\smartparagraph{Emerging policy workloads.}
Recent broadband policy initiatives impose measurement workloads that are fundamentally longitudinal in nature. 
A retrospective audit of the Connect America Fund (CAF) program showed that the absence of timely measurement and intervention undermined the effectiveness of a \$10B federal investment~\cite{caf-study}. The BEAD program further raises the stakes by committing over \$42B, requiring robust pre-disbursement baselines and ongoing assessment of how available broadband plans evolve as funds are deployed~\cite{bead-program}. In parallel, state and local affordability initiatives---including low-cost plan mandates~\cite{nyc-low-cost-plan,jcots} and competition-driven regulatory efforts~\cite{cpuc-competition-report}---require repeated assessment of pricing and market conditions to evaluate whether policy interventions translate into meaningful access outcomes. Collectively, these workloads require sustained, longitudinal, address-level visibility into broadband availability, affordability, and quality rather than one-time or static measurements.

\smartparagraph{Limitations of existing regulatory datasets.}
Meeting these longitudinal requirements cannot be achieved using existing regulatory broadband datasets alone. Historical mechanisms such as the Federal Communications Commission (FCC) Form~477~\cite{fcc-form477} relied on coarse, self-reported availability data and have been shown to substantially overstate coverage~\cite{form477-inaccuracy}. While more recent efforts, including the Broadband Data Collection (BDC)~\cite{fcc-bdc} and National Broadband Map (NBM)~\cite{fcc-nbm}, improve spatial resolution by collecting address-level availability and advertised speeds, they continue to depend primarily on provider self-reporting and provide limited visibility into pricing, eligibility constraints, and how broadband plans change over time~\cite{nbm-comcast-misreporting,nbm-overstating-coverage,isps-fined-for-not-reporting}. As a result, existing regulatory datasets cannot support longitudinal analysis of the broadband plans available to consumers at specific addresses.

\smartparagraph{Why query broadband plans directly from ISPs?} 
To obtain this missing visibility, policy-grade analysis must independently observe broadband plans as they are presented to consumers over time. ISP BATs provide an address-specific view of the broadband services available for purchase at a given location, including availability, advertised performance, pricing, and eligibility constraints. Systematic, longitudinal querying of BATs enables the construction of independent, address-level broadband plan datasets that can be refreshed and compared across a large and diverse set of providers, directly shaping the system design constraints for platforms intended to support emerging policy workloads.

\begin{figure}[t]
    \centering
   \begin{subfigure}[t]{0.47\linewidth}
        \centering
        \includegraphics[width=\linewidth]{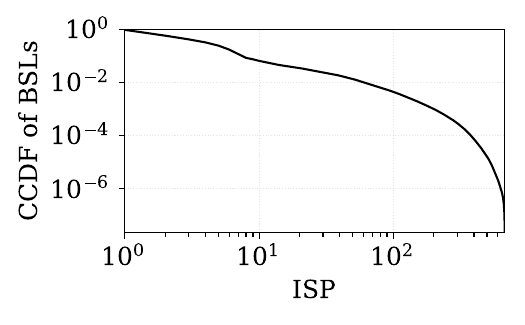}
        \caption{How many ISPs?}
        \label{fig:cdf-pew-national}
    \end{subfigure}
    \begin{subfigure}[t]{0.47\linewidth}
        \centering
        \includegraphics[width=\linewidth]{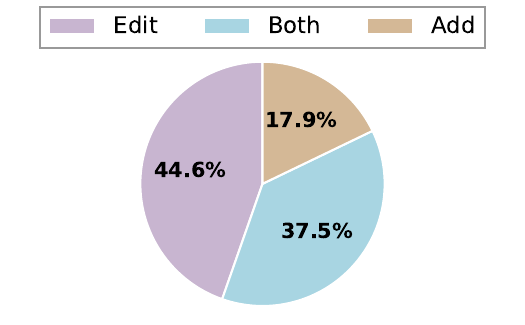}
        \caption{How dynamic are BATs?}
        \label{fig:pie-change-types}
    \end{subfigure}
 
    \caption{
    Empirical pressures imposed by policy-grade broadband measurement. (a)~Heavy-tailed provider landscape: cumulative fraction of broadband serviceable locations (BSLs) covered as the system supports more ISPs (log-scaled x-axis). (b)~BAT churn: breakdown of observed interface updates by whether they require adding new interaction states, editing existing ones, or both (56 updates over eight months across 64 longitudinally monitored ISPs).
    }
    \label{fig:isp-churn}
    \vspace*{-0.1in}
\end{figure}

\subsection{System Requirements}
\label{ssec:system-requirements}

Policy-grade, longitudinal broadband measurement raises two basic questions that drive system design: (1)~how many ISPs must a querying system support? and (2)~how dynamic are ISPs' BATs in practice?

\smartparagraph{How many ISPs must we support?}
National-scale policy workloads require covering a heavy-tailed provider ecosystem. Figure~\ref{fig:cdf-pew-national} shows the cumulative fraction of BEAD serviceable locations (BSLs) covered as a function of the number of serving ISPs. The top-10 providers, supported by BQT, cover roughly 80\% of BSLs, leaving a long tail of providers with small footprints (including 53 ISPs that serve only one BSL). One could argue that 80\% coverage is sufficient; however, from a policymaking perspective, this tail is critical: small, regional, and rural providers disproportionately correspond to underserved and underrepresented communities, so failing to support them systematically biases measurement away from the populations that programs like BEAD target. In all, establishing baselines and longitudinal tracking for programs like BEAD requires supporting \emph{hundreds} of providers.

\smartparagraph{How dynamic are BATs?} BATs evolve frequently and independently, and even minor UI changes can disrupt automated querying. To quantify this churn, we operate a longitudinal measurement pipeline over eight months. For each ISP, we repeatedly query a fixed set of ten curated addresses selected to elicit distinct terminal outcomes (e.g., serviceable vs.\ non-serviceable, or different plan pages), issuing queries once per week for the first five months and every two hours for the final three months. We define an \emph{interface update} as any instance in which repeated queries for the same address yield different observable interaction outcomes across successive measurements; Section~\ref{sec:evaluation} details the methodology. 

Across the 64 ISP BATs we monitored, we observed persistent, workflow-relevant evolution throughout the observation window, measuring a total of 56 interface changes, with 35 ISPs exhibiting at least one change. As shown in Figure~\ref{fig:pie-change-types}, these updates are often structural rather than purely cosmetic: they change the required user actions within an existing stage (e.g., an additional click to view plans), insert a new stage (e.g., a cookie popup), or combine both. While this count may not appear large in isolation, these are exactly the kinds of changes that break automated querying and necessitate interventions from technical experts to restore execution. At the policy scale (hundreds of ISPs over months to years), this cadence translates into recurring interventions that are untenable for non-expert stakeholders without dedicated engineering support.

\smartparagraph{Implications for system design.}
Taken together, these observations imply that policy-grade querying systems must simultaneously provide \textbf{\textit{extensibility}} to cover a heavy-tailed ISP landscape (hundreds of providers for BEAD-scale evaluation), \textbf{\textit{robustness}} to frequent, provider-specific BAT evolution (providers change independently and on different schedules), and \textbf{\textit{low technical overhead}} so authoring and maintenance remain tractable for non-expert stakeholders. Meeting all three is necessary to sustain long-running, address-level broadband measurement at the policy scale. These requirements point to an abstraction that \emph{disaggregates} intent (what to collect) from execution (how to navigate), represents interactions as a state space with multiple admissible paths, and makes interface evolution a localized specification update rather than an end-to-end workflow rewrite.

\subsection{Why BQT Falls Short}
\label{ssec:why-bqt-fails}

We now assess existing broadband plan querying systems through the lens of these abstraction requirements. We focus on BQT as a representative system as it exemplifies the workflow-centric approach adopted by prior broadband measurement efforts~\cite{bat,decoding-the-divide,markup-study} and has supported several policy analyses~\cite{decoding-the-divide,caf-study,bst,amicus_curiae}.

\smartparagraph{Workflow-centric abstraction.}
As shown in Figure~\ref{fig:BQT-vs-BQT+-system}, BQT represents each ISP BAT as a deterministic, end-to-end workflow implemented in low-level execution code (e.g., Python). This single artifact simultaneously encodes both querying intent (what attributes to extract) and execution strategy (the required interaction sequence), effectively committing the specification to one expected execution trace. This abstraction aligned with the operating regimes for which BQT was originally designed. Prior deployments targeted finite measurement campaigns, a limited set of major ISPs, and bounded time horizons~\cite{caf-study,decoding-the-divide,bst}, under which interface evolution was limited and manual workflow maintenance was feasible.

\smartparagraph{Mismatch under policy-scale conditions.}
Under the empirical conditions in Section~\ref{ssec:system-requirements}---a heavy-tailed provider ecosystem and frequent, provider-specific BAT evolution---this workflow-centric abstraction becomes brittle and expensive to maintain. Because interaction structure is collapsed into a single prescribed path, alternate but valid interface paths (or inserted/reordered steps) manifest as failures rather than admissible behavior. And because intent is expressed through imperative control flow, even localized interface changes tend to propagate into dispersed workflow edits, limiting reuse across ISPs and causing maintenance overhead to scale with coverage and time. 

These limitations reflect a fundamental consequence of the workflow-centric abstraction rather than shortcomings in engineering effort. Supporting policy-grade broadband measurement, therefore, requires a different abstraction—one that decouples querying intent from execution behavior and treats interface evolution as a modification of the interaction space. We introduce such an abstraction in the next section.

\section{Abstraction and System Design}
\label{sec:design}

This section introduces the proposed abstraction that addresses the requirements articulated in Section~\ref{ssec:system-requirements}. By representing querying intent independently of execution, the abstraction enables effective disaggregation. This disaggregation, in turn, makes robustness, extensibility, and low technical overhead inherent structural properties of the system.

\vspace{-\intextsep}
\subsection{Proposed Abstraction}
\label{ssec:abstraction}

\sys represents each ISP interaction as an \textbf{\emph{abstract NFA}} that captures the interaction structure required to obtain policy-relevant broadband attributes. The abstraction defines observable interface states and permissible user actions, while omitting execution order, control flow, retries, and recovery logic. At runtime, system mechanisms traverse this interaction space and adapt execution as interfaces evolve, *typically requiring only localized updates to state specifications (rather than end-to-end workflow rewrites). By decoupling interaction structure from execution behavior, the abstraction admits multiple admissible interaction paths, localizes the effects of interface changes, and enables reuse across providers at scale.

\smartparagraph{NFA-based representation of interaction structure.}
Formally, \sys models each ISP consumer-facing interface as an NFA defined by a tuple $(Q, \Sigma, \delta, q_0, F)$, where $Q$ denotes a finite set of interaction states, $\Sigma$ denotes a set of user actions, $\delta: Q \times \Sigma \rightarrow 2^Q$ defines a nondeterministic transition relation, $q_0 \in Q$ is the initial state, and $F \subseteq Q$ denotes terminal states~\cite{finite-automata}. The nondeterministic transition relation allows multiple successor states for a given state--action pair, explicitly encoding the presence of multiple admissible interaction paths. In \emph{BQT+}, $\delta$ captures which state--action pairs are admissible; the realized successor state is resolved at runtime by re-observing the page and matching detectors after executing the action (rather than being predicted statically).

In \emph{BQT+}, states correspond to observable interaction conditions exposed by the ISP interface, such as address entry, service qualification, or plan presentation. Actions correspond to user-initiated interactions that advance the interface. The NFA therefore specifies which state--action transitions constitute valid progress toward a querying objective, while intentionally avoiding any commitment to a particular execution trace or traversal order beyond constraining execution to admissible actions.

\smartparagraph{Nondeterminism as a first-order requirement.}
Empirical results in Section~\ref{ssec:system-requirements} show that ISP interfaces evolve by inserting intermediate steps, reordering interactions, and introducing conditional validation paths. These changes alter the set of valid interaction sequences over time, rather than preserving a single stable execution trace. A nondeterministic abstraction directly reflects this reality by representing the interaction space as a set of admissible paths rather than a single prescribed sequence.

Under this abstraction, interface evolution modifies the reachable structure of the interaction space by adding or altering states and transitions. Execution adapts by selecting alternative traversal paths at runtime, while the specification often requires only localized updates. This behavior supports sustained longitudinal operation under continuous interface change and directly addresses the empirical pressures identified in Section~\ref{ssec:system-requirements}.

\smartparagraph{Position in the workflow design space.}
Different workflow representations impose structural constraints that determine whether disaggregation is achievable. Deterministic finite automata~\cite{finite-automata} encode a single successor per state--action pair, forcing the specification to commit to one expected successor even when interfaces admit multiple valid outcomes, and reintroducing the fragility observed in workflow-centric systems~\cite{workflow-dfa}. Directed acyclic graphs (DAGs)~\cite{dag-workflows} permit branching but forbid cycles, preventing representation of retries, repeated validation, and iterative interaction states that arise in evolving ISP interfaces. For instance, dismissing or rejecting a cookie-consent pop-up may return the user to the same underlying state, forming a cyclic interaction pattern that DAG-based workflows cannot express. Petri nets~\cite{petri-nets} and BPMN-style models~\cite{bpmn} encode detailed execution semantics alongside interaction structure, increasing specification complexity and obstructing the separation of querying intent from execution behavior. Markov decision processes~\cite{mdp} and reinforcement-learning-based models~\cite{rl-web-agents} replace explicit interaction structure with learned policies, eliminating a stable, reusable representation of the interaction space and preventing reuse across providers. 

In contrast, an abstract NFA occupies a precise and minimal point in this design space. It captures evolving interaction structure, admits cycles and alternative paths, and remains finite, declarative, and execution-agnostic. These properties align directly with the abstraction requirements derived from policy-grade measurement workloads.

\subsection{Abstraction-Enabled Disaggregation}
\label{ssec:disaggregation}
The NFA abstraction enables disaggregation by separating declarative interaction semantics from procedural execution behavior. This separation defines a stable specification boundary: intent describes the interaction space that must be supported, while execution adapts dynamically within that space as interfaces evolve.

\smartparagraph{Intent--execution separation.}
In \emph{BQT+}, querying intent specifies the semantics of interaction rather than a sequence of actions. Intent defines which interface states are relevant, how those states are recognized, which user actions are admissible in each state, and which states constitute successful completion with associated data extraction. Execution behavior determines how the system traverses the interface at runtime, including action selection, traversal order, retries, and timing. The NFA encodes intent exclusively as interaction semantics, while runtime mechanisms implement execution by operating over the interaction space exposed by the NFA. This division assigns stability to intent and adaptability to execution: interface evolution changes how execution proceeds, not what intent means.

\smartparagraph{Intent as declarative interaction structure.}
The NFA represents querying intent as a finite interaction structure rather than as a prescribed execution trace. The initial state encodes entry conditions, terminal states represent successful completion, and transitions specify which actions are valid when a particular interface state is observed. The specification declaratively defines how the system should respond to any recognized state in order to make progress toward the querying objective, without encoding control flow, sequencing decisions, or recovery logic. By expressing intent as an interaction structure, the abstraction makes intent explicit, finite, and reusable across providers, while keeping execution strategies external to the specification.

\smartparagraph{Execution and interface evolution.}
Execution operates as a runtime traversal over the interaction space defined by the NFA. At each step, the system observes the current interface state, maps it to one or more NFA states, and selects an admissible action based on runtime conditions. Traversal decisions, recovery behavior, and retries depend on execution logic rather than static specification. Interface evolution, i.e., changes to an ISP's BAT, appears as a structural modification of the interaction space: new interface steps introduce additional states or transitions, reordered interactions alter transition relationships, and conditional validation paths expand branching structure. These changes update the NFA graph directly, while execution continues to traverse the updated space using the same runtime mechanisms. As a result, interface evolution modifies reachable structure rather than invalidating intent.

\smartparagraph{Emergent systems properties.}
By localizing interaction structure and decoupling it from execution, the system remains robust under routine interface evolution: changes affect only the relevant states and transitions, while querying intent remains valid and execution adapts through traversal rather than failure. The same structural representation enables extensibility across a large and heterogeneous ISP landscape by allowing shared interaction components to be reused and provider-specific behavior to be incorporated incrementally, without duplicating execution logic. At the same time, declarative intent specifications remain finite and stable over time, with maintenance effort confined to localized structural updates, keeping technical overhead tractable under sustained, longitudinal operation.

\smartparagraph{Key takeaway.}
This section shows that modeling query intent as an abstract NFA decouples intent from execution and enables disaggregation, making robustness and extensibility low-overhead properties rather than add-ons. This bridges policy-driven data requirements and sustained operation under continuous interface change.

\section{Implementation}
\label{sec:implementation}

\subsection{Intent Specification Interfaces}
\label{ssec:intent-specification}
\sys represents querying intent as a dictionary-based specification that instantiates an abstract NFA over an ISP's BAT. Each specification defines a finite set of abstract states, associates each state with one or more \emph{state detectors} that identify when the state is active based on observable interface cues, and enumerates a set of admissible \emph{actions} that may be executed in that state using a fixed action API. Each specification designates an initial state and one or more terminal \emph{outcome} states; terminal states may be annotated with extraction directives when policy-relevant broadband attributes are visible, and otherwise record non-plan outcomes (e.g., \emph{no service}).

\smartparagraph{Authoring intent specifications.}
\sys supports two interfaces for authoring intent specifications: a \emph{manual} interface and an \emph{agentic} interface. Both interfaces capture the same underlying interaction intent but allow users to express that intent at different levels of abstraction.

In the manual interface, users explicitly construct the specification by defining abstract states, specifying state detectors using observable interface cues, and enumerating admissible actions from the fixed action API. This process is typically guided by user-driven exploration of an ISP BAT. As the user navigates the interface (e.g., AT\&T's BAT), a screen recording is captured and processed to extract metadata that maps visible text elements to screen coordinates. Based on this recording, users identify interaction states (e.g., \texttt{ADDRESS\_BAR}) and define admissible actions that advance the interaction.

In the agentic interface, users instead provide a high-level natural language description of the intended interaction (e.g., ``enter the address and check availability''). An LLM-powered web actor analyzes the interface and synthesizes an intent specification by inferring abstract states, generating state detectors from observable cues, and translating natural language instructions into admissible actions drawn from the same fixed action API. Because intent is expressed more declaratively, the resulting abstract NFA may differ in state granularity or action decomposition from a manually authored one, while still capturing the same interaction semantics.
\vspace{-\intextsep}

\begin{figure*}[t]
   \centering
   \includegraphics[width=\linewidth,trim=0 0 0 20pt,clip]{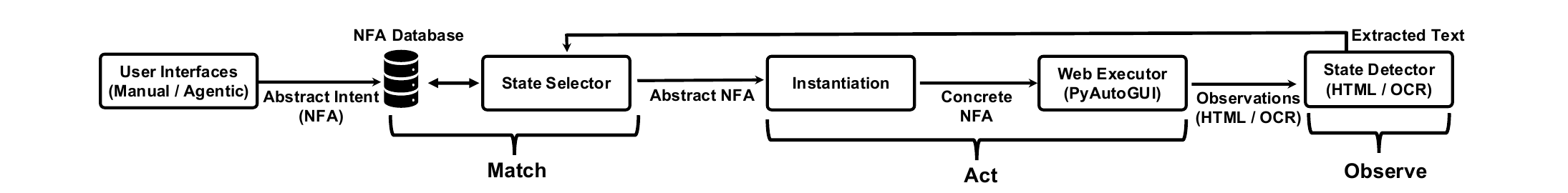}
   \vspace{-20pt}  
   \caption{\sys executes queries by compiling manual/agentic-authored abstract NFA intent into a concrete NFA and traversing it with an observe--match--act loop until a terminal state triggers extraction.}
   \label{fig:execution-loop}
\end{figure*}

\subsection{Abstract-to-Concrete NFA Instantiation}
\label{ssec:nfa-instantiation}

Intent specifications authored via either interface are not executed directly. Instead, \sys instantiates each abstract NFA into a \emph{concrete NFA} by binding abstract states to interface-observable conditions and abstract actions to executable interaction handlers. This instantiation step serves as a compilation and normalization phase that produces a uniform executable representation. Although the abstract NFAs produced by the manual and agentic interfaces may differ---reflecting differences in abstraction level or authoring style---instantiation reconciles these differences whenever they encode the same interaction semantics. As a result, both interfaces compile to an equivalent concrete NFA representation, which is the sole input to the runtime execution engine.

\smartparagraph{State instantiation.}
State instantiation defines how abstract states are recognized on rendered webpages. Each abstract state is associated with one or more \emph{state detectors}, which are predicates evaluated over observable interface cues such as visible text, structural markers, or the presence of specific UI elements. Multiple detectors may be bound to the same state to accommodate interface variation.

In the manual workflow, state detectors are derived from user-guided exploration of the ISP BAT. From the captured screen recording, users identify interaction states and specify detectors that characterize each state. For example, in AT\&T's BAT, an abstract state corresponding to the address entry page is instantiated using detectors that match phrases such as ``Enter your address'' and the presence of an address input field.

In the agentic workflow, state detectors are synthesized automatically using an LLM tool that analyzes rendered pages and DOM structure. While the agent-authored abstract NFA may introduce different state boundaries or levels of granularity, instantiation binds abstract states that correspond to the same interface-observable condition to a common concrete state.

\smartparagraph{Action instantiation.}
Action instantiation defines how abstract actions are realized as executable interaction handlers. Each abstract action is bound to a reusable \emph{action handler} selected from a fixed action API. Action handlers implement logical user interactions---such as clicking a UI element, entering text, submitting a form, or waiting for a condition---without encoding sequencing or recovery logic.

In the manual workflow, action metadata is derived from the same screen recording used for state instantiation. Users explicitly identify interface elements and specify concrete actions such as \texttt{typewrite(address)} or \texttt{click(Check Availability)}, which are realized using PyAutoGUI.

In the agentic workflow, an LLM tool translates natural language action descriptions into parameterized calls to the same action API. Although agent-authored abstract actions may be coarser or finer-grained than manually specified ones, instantiation normalizes these differences by binding them to equivalent concrete action handlers. Consequently, both workflows compile to an equivalent concrete NFA representation.
\vspace{-\intextsep}

\subsection{Observe--Match--Act Execution}
\label{ssec:runtime-execution}

\sys executes queries by traversing the concrete NFA using a unified observe--match--act control loop. Because runtime execution operates exclusively on the concrete NFA, execution semantics are identical regardless of how the abstract intent was authored. Figure~\ref{fig:execution-loop} presents \emph{BQT+}'s runtime loop.

\smartparagraph{Execution loop overview.}
Execution begins at the initial state of the concrete NFA. At each iteration, the runtime observes the rendered ISP BAT, matches the observation against concrete state detectors to identify active states, selects an admissible action associated with one of those states, and executes it. Execution completes upon reaching a terminal state, at which point any extraction directives associated with that terminal state are invoked. Execution halts unsuccessfully when no state matches the observed interface or when no admissible action remains.

\smartparagraph{State matching and progression.}
At each iteration, the runtime matches the observed interface against the concrete NFA to identify active states. Matching may yield zero, one, or multiple candidate states. When multiple states match, the runtime treats this as a \emph{mapping conflict} and refines the match before selecting an action.

Such conflicts arise because state detectors may overlap across different pages in an ISP workflow. To resolve conflicts, the runtime re-evaluates matching using alternative observation modalities, including Optical Character Recognition (OCR)-based detection that restricts matching to text visibly rendered on the page. If refinement fails, execution pauses and records the ambiguity for offline resolution. If no state matches the observed interface, the runtime treats the intent specification as \emph{underspecified}. The unmatched interface is recorded and may be used to extend the intent specification, either manually or via the agentic interface.

\smartparagraph{Action selection and execution.}
For a matched state, the runtime selects and executes a single admissible action, treating each action as an atomic unit of interaction. After executing an action, the runtime immediately re-observes the interface and re-matches states, rather than assuming a deterministic successor.

\smartparagraph{Progress, termination, and reuse.}
Execution proceeds by repeatedly applying the observe--match--act cycle until a terminal state is reached. \sys amortizes execution cost by caching instantiated states, bound actions, and execution artifacts. Because normalization occurs during abstract-to-concrete instantiation, cached concrete NFAs remain reusable across executions, addresses, and interface variants, reducing marginal engineering and execution cost as coverage expands.

\section{\sys Evaluation}
\label{sec:evaluation}
This section first assesses how \emph{BQT+}'s NFA abstraction enables extensibility and robustness while reducing the technical debt of operating \sys for large-scale policy workloads. It then demonstrates how agentic authoring, enabled by the NFA abstraction, improves accessibility for non-technical users.

\subsection{Experimental Setup}
\label{ssec:eval-framework}

\smartparagraph{Datasets.}
We develop a continuous measurement pipeline in which \sys queries, at regular intervals, a fixed set of ten curated street addresses per supported ISP, selected to elicit diverse interaction states and terminal outcomes in the ISP's BAT (e.g., non-serviceable, serviceable, and active-service addresses). Across ISPs, ten addresses capture most outcome diversity. The pipeline began operation in June 2025, initially running weekly, and later increasing to every two hours as coverage expanded. We base address selection on the motivating case studies in Section~\ref{sec:background} and reuse the same address set per ISP to ensure longitudinal consistency.

For each execution, the pipeline records the query \emph{hit rate}, defined as the fraction of address queries that reach a terminal outcome state. When hit rates drop, we inspect executions and update the intent specification. Interface evolution arises from new interaction states, changes that invalidate existing detectors or actions, or both. Each change triggers a targeted intervention, such as adding or editing an NFA state. We log every intervention, including the number of affected states and the modification type, to quantify the effort required for sustained operation.

Using this pipeline, we construct two datasets: (i)~a \textit{longitudinal dataset} spanning 64 ISPs, queried repeatedly over weeks to months, and (ii)~a \textit{coverage snapshot dataset} covering 100 ISPs queried over a single day, including 36 newly added within a couple of days. For both datasets, we log execution outcomes, hit rates, and all interface-driven updates to manually authored NFA specifications. We use the coverage snapshot dataset to evaluate extensibility (onboarding effort) and the longitudinal dataset to evaluate robustness under interface churn; unless noted otherwise, technical-debt analyses consider states observed across both datasets.

\smartparagraph{Baselines.}
We exclude the original BQT system from direct comparison because it does not enable a fair apples-to-apples evaluation: it targets a different ISP set, relies on static hand-written workflows, and lacks support for interface evolution. Instead, we use a \emph{BQT (Proxy)} baseline (simply referred to as BQT unless specified otherwise), which serves as a realistic proxy for BQT-style static deterministic workflows. This baseline consists of the concrete Python implementation obtained by fully materializing the abstract NFA specification into explicit control flow and state checks. By removing intent--execution decoupling while preserving identical semantics, this baseline enables a direct comparison of specification and maintenance overhead, with implications for both extensibility and robustness.

\smartparagraph{Metrics.}
We quantify specification overhead using \emph{Logical Lines of Specification (LLoS)} for \sys and \emph{Logical Lines of Code (LLoC)} for the \emph{BQT} baseline. We define LLoS as an abstraction-aware measure of intent complexity: for each NFA state, we count one logical line for the state identifier and one logical line for each associated action API call. We define LLoC~\cite{lloc} as the size of the corresponding imperative realization, measured as the number of executable Python statements obtained by fully materializing the same abstract specification into explicit control flow, state checks, and action sequencing. We also quantify authoring overhead across \emph{BQT+}'s manual and agentic interfaces using LLoS and \emph{Logical Specified Characters (LSC)}. LSC measures user typing effort and is defined as the total number of characters explicitly provided by the user (including whitespace and punctuation) as authoring input: for the manual interface, the serialized intent specification; for the agentic interface, the natural-language instruction(s) provided by the user. We exclude any system-generated text or templates. We primarily rely on these metrics and their derivatives to quantify extensibility, robustness, and technical debt.

\subsection{System Properties in Practice}
\label{ssec:eval-results}

This section reports results for the three system attributes defined in Section~\ref{ssec:eval-framework}: extensibility, robustness, and technical debt. We use the datasets and metrics introduced in Section~\ref{ssec:eval-framework} and focus exclusively on empirical results.

\smartparagraph{Extensibility.}
We evaluate extensibility using the \textit{coverage snapshot dataset} by measuring the specification effort required to onboard a new ISP. As a structural baseline, newly added ISPs require between 3 and 28 NFA states, with a median of 6 states, reflecting substantial diversity in interaction workflows. Table~\ref{tab:metrics-add-update-isps} shows summary statistics for states, Logical Lines of Specification (LLoS), and LLoC required to add a new ISP to \emph{BQT+}, measured at onboarding time. The statistics show that most ISPs require modest specification effort despite variability in workflow structure, with a small tail corresponding to particularly complex BATs containing a large number of intermediate states to extract comprehensive plan data (e.g., Xfinity and Cox Communications). These results show that intent--execution separation bounds onboarding effort as ISP coverage expands.

\begin{table}[h]
  \centering
  \caption{Extensibility (coverage snapshot): per-ISP onboarding effort; Robustness (longitudinal): per-intervention maintenance effort. We report states, LLoS, and the compiled BQT (Proxy) size (LLoC). \textit{Total} aggregates cumulative effort.}
  \label{tab:metrics-add-update-isps}
  \resizebox{\linewidth}{!}{%
  \begin{tabular}{l|cccccc}
    & \multicolumn{3}{c}{\textbf{Extensibility}} & \multicolumn{3}{c}{\textbf{Robustness}} \\
      & \multicolumn{2}{c}{\textbf{\sys}} & \textbf{BQT} & \multicolumn{2}{c}{\textbf{\sys}} & \textbf{BQT}\\
    & \textbf{States} & \textbf{LLoS}  & \textbf{LLoC} & \textbf{States} & \textbf{LLoS} & \textbf{LLoC} \\
    \hline
    Min & 3 & 8 & 725 & 0 & 0 & 0 \\
    25$^{th}$ percentile  & 5 & 15 & 779 & 1 & 2 & 17 \\
    Median & 6 & 19  & 805 & 2 & 5 & 33\\
    75$^{th}$ percentile & 8 & 31 & 855 & 5 & 14 & 94\\
    Max  & 28 & 88 & 1225 & 18 & 40 & 437 \\
    Average & 7.4 & 24.8 & 842.4 & 3.7 & 10.1 & 79.2 \\
    \hline
    Total & 723 & 2,420 & 81,717& 207 & 565 & 4,437 \\ 
  \end{tabular}%
}\\
\end{table}

\smartparagraph{Robustness.}
We evaluate robustness using the \textit{longitudinal dataset} (64 ISPs), in which we observe 56 total interface-driven interventions. For each intervention, we measure the number of NFA states added or edited and the resulting change in LLoS. Table~\ref{tab:metrics-add-update-isps} shows a summary of (i)~the number of states affected per intervention, (ii)~the corresponding change in LLoS per intervention, and (iii)~the implied change in the size of the compiled execution (LLoC) per intervention (BQT). The summary statistics show that most interventions affect only a small number of states and induce limited changes to the abstract specification.

Exceptions are when the ISP redesigns its interface, in which multiple states must be modified or added. For instance, following Astound Broadband's interface update, extracting all plan data requires individually clicking each Broadband Label button, necessitating changes to 18 states in a single intervention. Similarly, Cox Communications' major interface redesign required modifications in 11 states. However, when an ISP changes only the placement of existing UI elements without altering the interface structure, no NFA states need to be modified. In these cases---such as DayStarr Communications---updates are handled through modifications to element locations using the screen recording process. While interventions typically require modest changes in states and LLoS, the implied change in compiled execution size (LLoC) is substantially larger for the same interventions. This disparity demonstrates that interface evolution maps to localized intent updates rather than cascading workflow rewrites, and that robustness in \sys arises from intent localization through workflow decomposition.

\begin{figure}[t]
    \centering
    \begin{subfigure}[t]{0.48\linewidth}
       \centering
       \includegraphics[width=\linewidth]{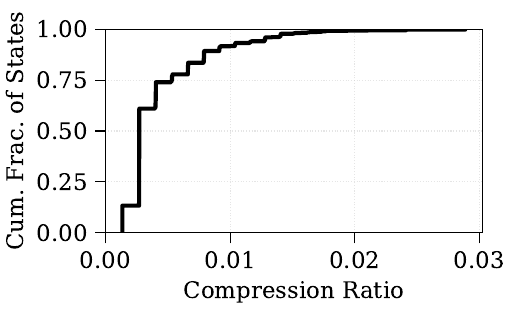}
       \caption{Compression.}
       \label{fig:compression-ratio}
    \end{subfigure}
    \hfill
    \begin{subfigure}[t]{0.48\linewidth}
       \centering
       \includegraphics[width=\linewidth]{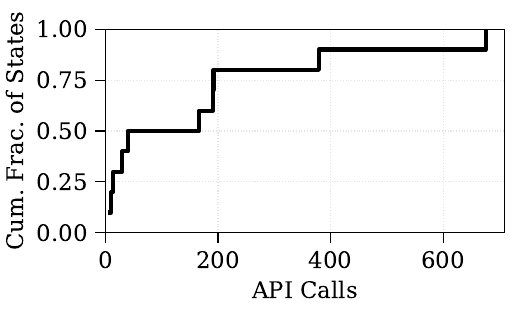}
       \caption{API usage.}
       \label{fig:api-usage}
    \end{subfigure}
    
    \caption{Technical-debt indicators: (a) per-state compression; (b) action API usage.}
    \label{fig:cdfs-compression-api-calls}
\end{figure}

\smartparagraph{Technical debt.}
We evaluate technical debt by measuring whether specification and cognitive complexity accumulates as coverage expands across ISPs and over time. To isolate abstraction-induced compression from onboarding or intervention effects, we compute, for each NFA state in the longitudinal dataset, the ratio between its abstract specification cost (LLoS contribution) and its concrete execution footprint (LLoC contribution). Figure~\ref{fig:compression-ratio} plots the CDF of this per-state compression ratio across all ISPs and time. The distribution concentrates at low ratios, showing that individual interaction states consistently replace substantially larger execution logic, limiting per-state cognitive overhead as the system scales.

We next analyze reuse of execution primitives. Figure~\ref{fig:api-usage} plots the frequency distribution of admissible action API calls aggregated across all NFA states and ISPs. The distribution exhibits a heavy-tailed shape, with a small set of core primitives (click, finalize, keypress, and typewrite) accounting for the majority of usage. This pattern indicates that supporting new ISPs primarily composes existing execution primitives rather than introducing new ones.

\begin{figure}[t]
    \centering
    \begin{subfigure}[t]{0.48\linewidth}
       \centering
       \includegraphics[width=\linewidth]{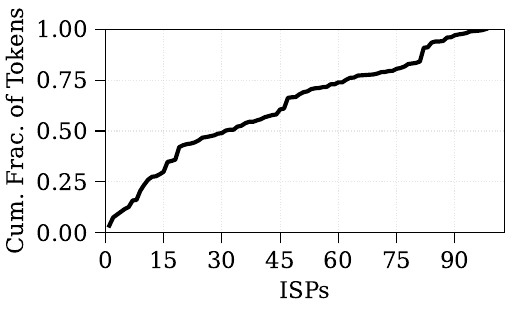}
       \caption{Token growth.}
       \label{fig:detector-vocab}
    \end{subfigure}
    \hfill
    \begin{subfigure}[t]{0.48\linewidth}
       \centering
       \includegraphics[width=\linewidth]{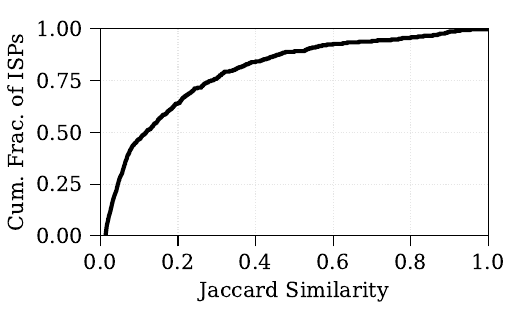}
       \caption{Similarity.}
       \label{fig:detector-similarity}
    \end{subfigure}
    
    \caption{Detector reuse: (a) token growth with ISP coverage; (b) pairwise Jaccard similarity between detectors.}
    \label{fig:cdfs-detector-reuse}
\end{figure}

Finally, we analyze reuse of state detector cues. Figure~\ref{fig:detector-vocab} plots the cumulative number of unique detector tokens as ISPs are added to the system. The resulting curve grows sublinearly with coverage, indicating that new ISPs rarely introduce novel detection concepts and instead reuse an existing vocabulary of interaction cues. Figure~\ref{fig:detector-similarity} reports the distribution of pairwise Jaccard similarity over tokenized detector strings, where most detectors exhibit substantial overlap with existing ones. Together, Figure~\ref{fig:cdfs-detector-reuse} shows that new ISPs reuse existing interaction cues rather than introducing novel detector concepts. Overall, these results demonstrate that \sys limits technical debt by compressing execution complexity at the level of individual interaction states and by promoting reuse of both action primitives and detector vocabulary as coverage scales.

\subsection{Suitability for LLM-Powered Authoring}
\label{ssec:agentic-eval}

We evaluate whether \emph{BQT+}'s intent abstraction supports LLM-powered authoring along two dimensions: (1)~does agentic authoring reduce specification overhead, and (2)~does synthesized intent preserve the intended executable behavior? The agentic interface is a recent addition to \sys and currently supports intent authoring for approximately 100 ISPs. We evaluate it using the coverage snapshot dataset via a controlled execution study with a single validation round and 10 curated addresses per ISP.

\begin{figure}[h]
    \centering
    \includegraphics[width=0.7\linewidth]{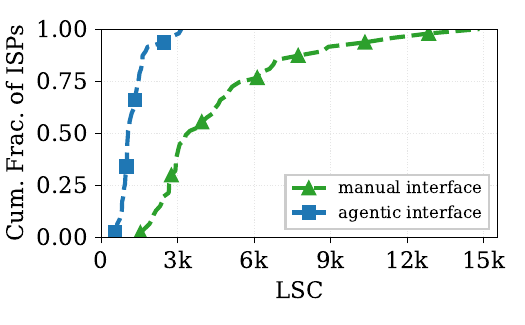}
    \vspace{-\intextsep}
    \caption{LSC needed to add a new ISP to \emph{BQT+}.}
    \label{fig:lsc-agentic-vs-manual}
\end{figure}

To answer the first question, we compare the authoring input size required by the manual and agentic interfaces. Figure~\ref{fig:lsc-agentic-vs-manual} reports the LSC for both interfaces. Agentic authoring consistently reduces LSC across ISPs, indicating that the abstraction exposes intent at a level that allows an LLM to synthesize executable specifications with less user-provided text.

To answer the second question, we evaluate execution fidelity. We select 50 ISPs supported by both interfaces and execute the same 10 curated addresses per ISP under identical runtime semantics. We treat the terminal outcome state reached by the manually authored specification as a reference label and measure the fraction of executions in which the agentic specification reaches the same terminal outcome state. In this evaluation, we observe 100\% agreement over the tested ISP--address pairs, indicating that agent-authored intent reproduces the manual specification's observable terminal outcomes in this controlled setting.

\section{\sys in Action}
\label{sec:studies}

This section introduces the policy-scale broadband measurement enabled by \emph{BQT+}. We present the methodology for each study, describe the results, and conclude by examining broader implications. 

\subsection{Empowering Legislative Actions}
\label{ssec:affordability}

\smartparagraph{Policy motivation.}
Several U.S. states are actively considering legislative tools to improve broadband affordability, including rate regulation and mandated low-cost plans. New York has already enacted a statewide affordability requirement, and other states are evaluating similar interventions. In this context, the Virginia General Assembly tasked the Joint Commission on Technology and Science (JCOTS)~\cite{jcots} with assessing broadband affordability in the Commonwealth and recommending concrete legislative actions. This charge requires addressing two coupled questions: (i)~what constitutes a defensible affordability benchmark for a low-cost broadband plan, and (ii)~to what extent existing market plans meet that benchmark in practice.

\smartparagraph{Data collection.}
To support this assessment, we use \sys to independently collect address-level broadband plan data across Virginia. We focus on ten representative localities---six counties (Pittsylvania, Halifax, Rockbridge, Loudoun, Fauquier, and Albemarle) and four cities (Portsmouth, Martinsville, Harrisonburg, and Richmond)---selected to capture Virginia's geographic, socioeconomic, and infrastructural diversity. Using census block groups (CBGs) as the unit of analysis, we sample at least 30 residential street addresses per CBG--ISP pair and query each address using \emph{BQT+}. The resulting dataset spans approximately 900 CBGs and 62{,}000 addresses across 10 providers and access types, including wired providers (Xfinity, Verizon, Cox Communications, Riverstreet Networks, Ting, and Lumos) and fixed wireless access (FWA) plans (AT\&T FWA, Verizon FWA, Riverstreet Networks FWA, and All Points Broadband FWA). For each CBG--ISP pair, we identify the lowest-priced generally available plan meeting the FCC's 100~Mbps minimum download threshold. Full dataset details appear in Appendix Table~\ref{tab:affordability-table}.

\begin{figure}[t!]
   \centering
   \includegraphics[width=\linewidth]{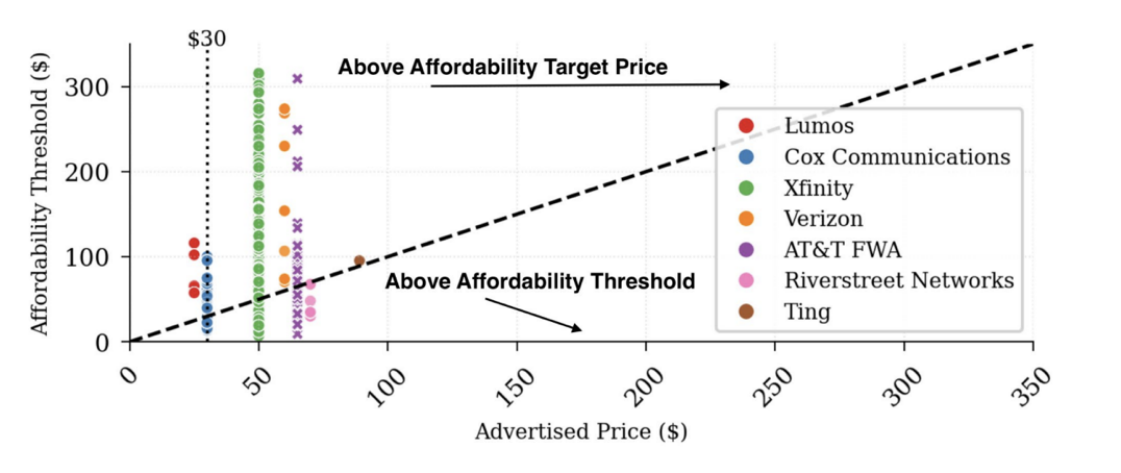}
   \caption{Broadband affordability frontier for Virginia.}
   \label{fig:affordability-frontier}
\end{figure}

\smartparagraph{Affordability benchmark.}
We define broadband affordability using a 2\% income threshold applied to the 20$^{th}$ percentile of disposable household income, consistent with the FCC's standards~\cite{income-threshold}. Using 2023 American Community Survey (ACS) data, we compute an affordability threshold for each CBG in Virginia. Under this criterion, a broadband plan priced at \$30/month is affordable for approximately 93\% of Virginia's population, whereas higher thresholds (e.g., \$50/month) leave roughly half of households with unaffordable options. Therefore, we establish \$30/month as a defensible statewide affordability benchmark for low-cost broadband.

\smartparagraph{Gap between benchmark and market plans.}
Figure~\ref{fig:affordability-frontier} summarizes the current state of broadband affordability in Virginia using an affordability frontier plot. Each point represents a CBG; the x-axis shows the advertised price of the available low-cost plan in that CBG, and the y-axis shows the CBG-specific affordability threshold. The vertical dashed line marks the \$30 price benchmark, and the diagonal line denotes the affordability frontier. Points above the affordability frontier indicate CBGs where at least one provider offers a plan priced below the affordability threshold, while points below the affordability frontier indicate CBGs where all available low-cost plans exceed the affordability benchmark.

Despite the fact that a \$30 benchmark would render broadband affordable for most Virginians, existing market plans systematically fail to meet this target. Statewide, only 0.45\% of CBGs have access to broadband plans priced at or below \$30/month---primarily offered by Cox Communications (\$30) and Lumos (\$25)---and approximately 41.2\% of CBGs lack access to any plan that meets the affordability threshold. Xfinity offers the lowest-priced qualifying plans in roughly 80\% of studied CBGs, yet these plans start at \$50/month.

\smartparagraph{Additional findings.}
Beyond the primary affordability gap, the data reveals additional structural barriers. Even when low-cost plans exist, they are often difficult to discover on ISP BATs or require contacting customer service, creating friction for low-income and digitally limited households. In some cases, plan visibility changes over time; for example, Xfinity did not initially surface low-cost plans on its plans page, but later updated its interface to display them. Moreover, increased provider competition does not reliably translate into lower low-cost plan prices, challenging common assumptions that market forces alone resolve affordability concerns. Finally, in contrast to common expectations~\cite{fwa-cheaper}, fixed wireless access fails to close the affordability gap: in approximately 90\% of CBGs where both wired and fixed wireless options are available, fixed wireless plans are \$10--\$60 more expensive while offering lower average speeds.

\smartparagraph{Policy implications.}
Based on this analysis, JCOTS recommended to the Virginia General Assembly in Fall~2025 that the state (i)~establish a statewide requirement that ISPs offer a 100~Mbps broadband plan priced at \$30/month, (ii)~mandate visibility and accessibility of low-cost plans on ISP websites, and (iii)~support targeted bridge programs to address residual affordability gaps. Critically, the committee also recommended independent, longitudinal data collection using tools such as \sys to sustain evidence as broadband markets and ISP interfaces evolve. This study demonstrates how robust, extensible data-generation systems can directly inform legislative decision-making by transforming heterogeneous consumer-facing ISP interfaces into policy-grade evidence.

\subsection{Enabling Intervention Infrastructure}
\label{ssec:bead-baseline}

\smartparagraph{Policy motivation.}
The \$42B BEAD program is the largest broadband infrastructure investment in U.S. history, yet experiences from prior programs, most notably CAF, show that infrastructure spending without independent, timely oversight can fail to produce affordable and adoptable service~\cite{caf-study}. In this context, the Pew Charitable Trusts~\cite{pew-charitable-trusts}---a nonpartisan policy think tank actively engaged in broadband policy---sought to determine whether it is empirically feasible to build an intervention infrastructure capable of tracking BEAD's progress and enabling early action when outcomes diverge from program goals. Doing so requires establishing a pre-disbursement baseline that characterizes broadband quality and affordability in BEAD-eligible locations, prior to any BEAD-induced changes.

\smartparagraph{Baseline data collection.}
To this end, we focus on BEAD-eligible CBGs across 64 ISPs and four representative states with substantial geographic, demographic, and partisan diversity: California, Michigan, Oklahoma, and Virginia~\cite{bead-baseline-study}. Using address-level querying, we collect advertised plan information across approximately 62{,}000 residential addresses spanning roughly 1{,}000 CBGs. We define BEAD-eligible CBGs as those in which at least 50\% of broadband serviceable locations (BSLs) are unserved or underserved, following National Telecommunications and Information Administration (NTIA) guidance. This threshold provides broad coverage of locations likely to be affected by BEAD while enabling a holistic assessment of baseline service conditions. Full dataset details appear in Appendix Table~\ref{tab:queried-BSLs}.

\smartparagraph{Baseline service conditions in BEAD-eligible areas.}
Table~\ref{tab:bead-baseline-50} summarizes baseline service conditions in BEAD-eligible CBGs using the 50\% eligibility threshold. We compute these statistics from affordability--availability frontier plots analogous to Figure~\ref{fig:affordability-frontier}, shown for all four states in Appendix Figure~\ref{fig:all_states_comparison_50}. As in Section~\ref{ssec:affordability}, we define affordability using an income-based threshold, treating broadband as unaffordable when subscription costs exceed 2\% of local household income. Across all four states, a substantial fraction of BEAD-eligible locations face persistent affordability challenges: between 60\% (California) and 77\% (Michigan) of CBGs only have access to plans priced above the affordability threshold. Availability gaps are pronounced as well: between 19.3\% (Michigan) and 50\% (Oklahoma) of BEAD-eligible CBGs lack access to a plan meeting the FCC's 100~Mbps download benchmark. Note that Virginia's analysis is limited because Brightspeed's BAT does not display plan-level pricing or speed tiers. Consequently, \sys extracts plan attributes for only 20.26\% of queried Virginia addresses, and Virginia's baseline statistics should be interpreted with caution.

\begin{table}[t]
  \centering
  \caption{Baseline service conditions in BEAD-eligible census block groups.}
  \label{tab:bead-baseline-50}
  \resizebox{.8\columnwidth}{!}{%
  \begin{tabular}{l|cc}
    & \textbf{ Price > FCC threshold} & \textbf{Speed < 100 Mbps} \\
    \hline
    CA & 60\% & 36\% \\
    MI  & 77\% & 19.3\% \\
    OK  & 74\% & 50\% \\
    VA  & 61\% & 0\% \\
  \end{tabular}%
  }
\end{table}

\smartparagraph{Affordability risk under constrained pricing oversight.}
Although BEAD targets improvements in availability and service quality, recent NTIA guidance limits states' ability to impose explicit price-related conditions on BEAD-funded deployments. As a result, infrastructure investment may expand coverage while failing to deliver affordable service to low-income households. Our baseline analysis shows that affordability challenges already prevail in BEAD-eligible areas prior to deployment, indicating that upgrades in network capacity alone may not translate into increased adoption or equity gains.

\smartparagraph{Policy implications.}
These findings underscore the need for independent, longitudinal tracking of both availability and affordability throughout BEAD implementation. By enabling scalable, sustained address-level data collection across heterogeneous and evolving ISP interfaces, \sys supports BEAD oversight at national scale and equips policy advocates, state broadband offices, and legislators with timely, actionable evidence for intervention.

\section{Limitations}
\label{sec:limitations}

\smartparagraph{Advertised vs. experienced service quality.} \sys only measures advertised availability and price, but does not measure actual, delivered performance. Previous work~\cite{bst} has highlighted the gaps between advertised and experienced service quality. Thus, our findings offer only an optimistic estimate of service quality for customers.

\smartparagraph{Queryable ISP interfaces.} 
Not all ISPs maintain a queryable web interface, and some ISPs do not display special qualifying offers through their primary interfaces. For example, low-cost plans that require consumers to visit a separate site and verify eligibility are outside of \emph{BQT+}'s scope. Some ISPs do not display available broadband plans at locations with active subscribers, limiting \emph{BQT+}’s ability to extract representative samples in regions with high adoption rates. 

\smartparagraph{CAPTCHAs.} 
\sys avoids triggering CAPTCHAs by mimicking human behavior, but does not solve them if triggered. \emph{BQT+}'s agentic interface relies on JavaScript injection for screen detection, which can trigger CAPTCHA mechanisms. In contrast, the manual interface does not trigger CAPTCHAs. 

\smartparagraph{Manual overhead for intent specification.}
When specifying intent, the user must iteratively experiment with different state--action sequences (in the manual interface) or prompts (in the agentic interface) to identify a workflow that executes without triggering bot-detection mechanisms.

\section{Related Work}
\label{sec:related_work}
Unlike generic web scraping, ISP BATs are interactive, stateful workflows that must be executed to reveal plans, and they evolve over time, necessitating a different approach.

\smartparagraph{Broadband measurement via ISP interfaces and other data sources.}
BQT enabled large-scale, address-level measurement by querying ISP interfaces and has been used in multiple policy studies (e.g., nationwide affordability analysis~\cite{decoding-the-divide} and CAF evaluation~\cite{caf-study}). Complementary efforts have derived broadband insights from adjacent artifacts and datasets, such as audits of FCC Broadband Labels~\cite{TPRC25_nutrition_labels} and analyses combining regulatory data with crowdsourced speed tests~\cite{imc_redIsSus}. Other work reverse-engineered BAT backend APIs to issue direct requests~\cite{bat}, yielding evidence of overstated coverage but potentially diverging from the consumer-facing BAT workflow. In contrast, \sys focuses on sustaining consumer-visible, longitudinal BAT querying under interface evolution.

\smartparagraph{Automation toolchains and scraping services.}
Browser automation frameworks ~\cite{pyautogui,seleniumbase, puppeteer,playwright}) provide actuation primitives, but scripts typically hard-code a single interaction trace and break under UI drift. Commercial scraping services~\cite{scraper-api,scrapingbee,scrapingdog,parsehub,octoparse,scrapy} help with request logistics, but not with maintaining evolving, provider-specific BAT workflows at policy scale. \sys builds on these tools by contributing an abstraction that localizes updates and enables alternate interaction paths at runtime.

\smartparagraph{LLM-based web agents.}
Recent multimodal web agents~\cite{gemini-computer-use,anthropic-computer-use,web-agent,web-voyager,reflexion,agent-q,netgent25,agent-s,agent-s2} make progress on generic website interaction, but they are a poor fit as the \emph{runtime} substrate for policy-grade BAT measurement. Our workload cannot tolerate either high error rates---each query must reliably traverse to one of a small set of terminal outcomes, and small action/state-recognition mistakes quickly collapse hit rates---or high per-query token/API cost from repeated perception--reasoning loops at scale~\cite{webarena,osworld,gpt-4v}. Instead, \sys uses an auditable NFA-based intent representation to constrain execution to admissible paths and localize updates under churn, while still allowing LLMs to assist with intent authoring.

\section{Conclusion}
\label{section:conclusion}

In this paper, we introduce a novel interaction-state abstraction that separates querying intent from execution. We  present \emph{BQT+}, an extensible and robust framework for large-scale broadband plan data collection that addresses fundamental limitations of existing broadband datasets and the original BQT architecture. By decoupling workflow intent from execution and modeling intent using an NFA abstraction, \sys lowers the threshold of adapting to ISP website changes and adding new ISPs. Using \emph{BQT+}, we establish a baseline for the BEAD program and analyze broadband affordability. Together, these datasets offer a deeper view of broadband disparities, thus informing equitable funding allocation and infrastructure investment. Overall, \sys provides a practical foundation for transparent and evidence-based broadband policymaking, as well as a case for longitudinal and scalable broadband data collection. \sys represents a critical step towards democratizing access to accurate broadband affordability data, empowering communities and policymakers to close Internet access gaps through open, verifiable, and data-driven approaches. 

\smartparagraph{Ethical considerations.} We collect data at street-address granularity and do not collect or analyze PII. Addresses are provided by Zillow and CostQuest Associates under a data use agreement and contain no resident-identifying information. We throttle queries, do not solve CAPTCHAs, do not create accounts or bypass paywalls, and limit measurements to consumer-facing qualification interfaces. To minimize load, we reuse a fixed set of ten addresses per ISP for longitudinal monitoring.

\label{endofbody}

\bibliographystyle{ACM-Reference-Format}
\bibliography{refs}

\appendix
\section{Appendix}

\subsection{Broadband Affordability Dataset}
\label{ssec:appendix-affordability-data}

Table~\ref{tab:affordability-table} summarizes ISP coverage, market structure, and low-cost plans across the 10 ISPs and 10 localities in this study, highlighting gaps between affordability benchmarks and current market broadband plans. Column 1 lists the ISPs. Column 2 reports the fraction of CBGs that the ISP serves, out of the total 897 CBGs in this study. Column 3 reports the competitive market conditions for the ISP, indicating the share of CBGs where the ISP operates as a monopoly, duopoly, or in competition with two or more providers (triopoly+). Column 4 indicates the fraction of CBGs where the ISP offers the low-cost plan. 

\begin{table*}[t]
  \centering
  \caption{Summary of ISP coverage, market structure, and low-cost plan characteristics.}
  \label{tab:affordability-table}
  \resizebox{.9\textwidth}{!}{%
  \begin{tabular}{l|cccc|ccc}
    & & \multicolumn{3}{c|}{\textbf{Market Structure (\%)}} & \multicolumn{3}{c}{\textbf{Low-Cost Plans}} \\
 & \textbf{Coverage (\%)} & \textbf{Monopoly} & \textbf{Duopoly} & \textbf{Triopoly+} & \textbf{(\%)} & \textbf{Speed (Mbps)} & \textbf{Price (\$)}\\
    \hline
    Xfinity & 80.5 & 3.2 & 8.4 & 68.8 & 81.4 & 300 & 50 \\
    AT\&T (FWA) & 84.8 & 0 & 6.1 & 78.7 & 7.37 & 90-300 & 65 \\
    Verizon (FWA) & 65.1 & 0.2 & 2.9 & 61.98 & 0 &  &  \\
    Verizon & 44.3 & 0 & 0.9 & 43.4 & 0.79 & 300 & 60-85 \\
    Cox & 9.0 & 0 & 0.7 & 8.4 & 9.18 & 100 & 30 \\
    Riverstreet Networks & 4.8 & 0 & 0.1 & 4.7 & 0.45 & 100 & 70 \\
    Riverstreet Networks (FWA) & 4.8 & 0 & 0.1 & 4.7 & 0 &  & \\
    Ting & 1.8 & 0 & 0 & 1.8 & 0.11 & 1000 & 89 \\
    Lumos & 1.1 & 0 & 0.1 & 1.0 & 0.68 & 300 & 25 \\
    All Points Broadband (FWA) & 0.7 & 0 & 0 & 0.7 & 0 &  &  \\
    \hline
    All ISPs & 100.0 & 3.4 & 27.1 & 69.5 &  & &  \\
  \end{tabular}%
  }%
\end{table*}

\subsection{BEAD Dataset}
\label{ssec:appendix-bead-dataset}

To establish a pre-disbursement baseline for the BEAD program within BEAD-eligible CBGs, we leverage \sys to query approximately 62,000 residential addresses. We classify \sys results into three categories: (1) Serviceable, (2) No Service and (3) Unknown. An address is classified as Serviceable if \sys is able to successfully navigate to the advertised plans page, and extract broadband plans. Within the Serviceable category, we distinguish between addresses for which broadband plans are displayed and those for which plans are not shown. Some ISPs (e.g., Brightspeed) do not display the broadband plans available at an address and instead prompt the customer to call a customer service representative for more details. When an ISP indicates that service is available at an address---even without listing specific broadband plans---we still classify the address as Serviceable. An address is classified as ``No Service'' if \sys can enter the address into the ISP's website and select a matching entry from the dropdown, but the ISP indicates that the address is outside its service area. Finally, an address is classified as ``Unknown'' if \sys cannot reach the plans page---either because the address is not listed in the ISP's dropdown menu and is therefore treated as invalid or when the ISP indicates potential service at the address, but does not explicitly confirm whether service is available. Table~\ref{tab:queried-BSLs} provides a detailed breakdown of \sys data collection across the four states, including the number of BSLs, CBGs and ISPs covered, within CBGs where at least 50\% of BSLs are BEAD-eligible. 

\begin{table*}[t]
  \scriptsize
  \centering
  \caption{Total number of queried BSLs, CBGs, and ISPs by state.}
  \label{tab:queried-BSLs}
  \resizebox{\textwidth}{!}{%
    \begin{tabular}{c|c c c c c c c}
      & \textbf{Total Addresses} &
      \textbf{CBGs} &
      \textbf{ISPs} &
      \textbf{Serviceable BSLs} &
      \textbf{Serviceable BSLs} &
      \textbf{No Service BSLs} &
      \textbf{Unknown BSLs} \\
      & & & & \textbf{(with plans)} & \textbf{(no plans)} & \\
      \hline
      CA & 20,135 & 454 & 19 & 52.68\% & 6.91\%  & 9.40\% & 31.01\% \\
      MI & 35,768 & 479 & 18 & 54.59\% & 0\%     & 9.49\% & 35.92\% \\
      OK & 2,488  & 60  & 11 & 33.52\% & 12.50\% & 11.01\% & 42.97\% \\
      VA & 3,969  & 58  & 16 & 20.26\% & 7.84\%  & 7.0\% &  64.90\% \\
      \hline
      Total & 62,360 & 1,051 & 64 &  &  &  \\
    \end{tabular}
  }
\end{table*}

\subsection{Affordability Frontiers (BEAD)}
\label{ssec:appendix-frontiers}
\begin{figure}[hbt!]
   \centering
   \begin{subfigure}[b]{0.48\columnwidth}
       \centering
       \includegraphics[width=\linewidth]{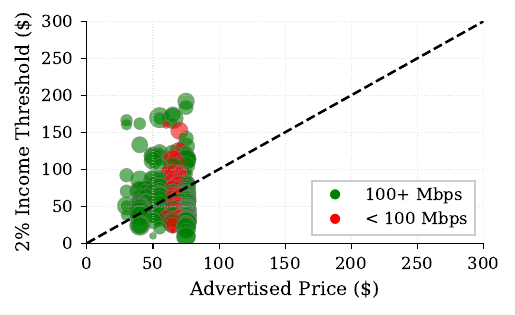}
       \caption{California}
   \end{subfigure}
   \hfill
   \begin{subfigure}[b]{0.48\columnwidth}
       \centering
       \includegraphics[width=\linewidth]{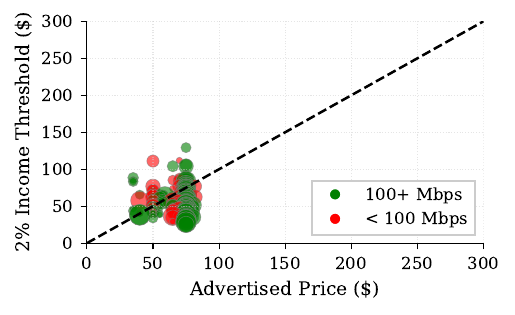}
       \caption{Michigan}
   \end{subfigure}

   \vspace{0.5em}

   \begin{subfigure}[b]{0.48\columnwidth}
       \centering
       \includegraphics[width=\linewidth]{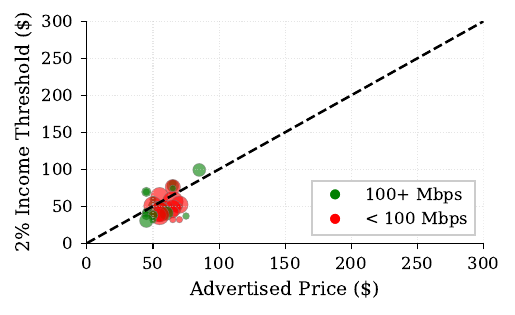}
       \caption{Oklahoma}
   \end{subfigure}
   \hfill
   \begin{subfigure}[b]{0.48\columnwidth}
       \centering
       \includegraphics[width=\linewidth]{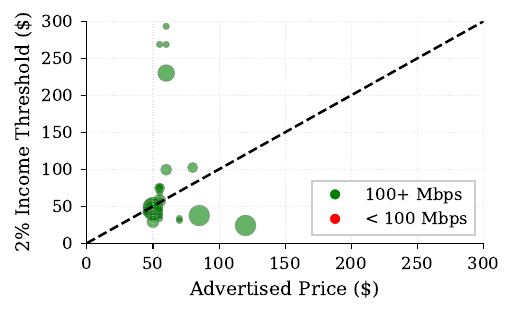}
       \caption{Virginia}
   \end{subfigure}

   \caption{Broadband affordability analysis across four states. Each dot represents a CBG, with green indicating representative plans $\geq$100 Mbps and red indicating representative plans \textless{}100~Mbps. Dot size reflects data coverage quality. Diagonal dashed lines indicate income affordability boundaries (dots below the diagonal exceed the 2\% income threshold).}
   \label{fig:all_states_comparison_50}
\end{figure}

Following \sys data collection, we compute a representative plan for each CBG. We first select, at each Broadband Service Location (BSL), the plan whose download speed is closest to 100~Mbps using regular (post-promotional) pricing, and then aggregate these selections to the CBG level by taking the median speed and the median price among BSLs offering that speed. Figure~\ref{fig:all_states_comparison_50} visually compares baseline service conditions across all four states using affordability frontier plots. In each subfigure, the representative broadband plan price for each CBG (horizontal axis) is plotted against the corresponding 2\% income affordability threshold (vertical axis). The diagonal line represents the affordability frontier---dots below this line indicate CBGs where the price of the representative plan exceeds what low-income households can afford. Green dots represent CBGs where the representative plan meets or exceeds the 100~Mbps download speed, while red dots indicate CBGs where the representative plan falls below 100~Mbps. Each dot corresponds to a single CBG. The size of the dot reflects the quality of data collection, with larger dots indicating CBGs where \sys successfully curated plans from a higher proportion of queried addresses. 

\end{sloppypar}
\end{document}